\documentclass[12pt]{iopart}
\usepackage{graphicx}
\expandafter\let\csname equation*\endcsname\relax
\expandafter\let\csname endequation*\endcsname\relax
\usepackage{amsmath}

\begin{document}

\title[Quantum sensing with tunable superconducting qubits]{Quantum sensing with tunable superconducting qubits: optimization and speed-up}

\author{S Danilin$^1$\footnote{Present address:
Oxford Quantum Circuits, Thames Valley Science Park, Shinfield, Reading
RG2 9LH United Kingdom}, N Nugent$^1$ and M Weides$^1$}

\address{$^1$ James Watt School of Engineering, University of Glasgow, G12 8QQ}
\ead{Martin.Weides@glasgow.ac.uk}

\begin{abstract}
Sensing and metrology are crucial in both fundamental science and practical applications. They meet the constant demand for precise data, enabling more dependable assessments of theoretical models' validity.  Sensors, now a common feature in many fields, play a vital role in applications like gravity imaging, geology, navigation, security, timekeeping, spectroscopy, chemistry, magnetometry, healthcare, and medicine. The advancements in quantum technologies have sparked interest in employing quantum systems as sensors, offering enhanced capabilities and new possibilities. This article describes the optimization of the quantum-enhanced sensing of magnetic fluxes with a Kitaev phase estimation algorithm based on frequency tunable transmon qubits. It provides the optimal flux biasing point for sensors with different qubit transition frequencies and gives an estimation of decoherence rates and achievable sensitivity. The use of $2$- and $3$-qubit entangled states are compared in simulation with the single-qubit case. The flux sensing accuracy reaches $10^{-8}\cdot\Phi_0$ and scales inversely with time, which proves the speed-up of sensing with high ultimate accuracy.
\end{abstract}

\maketitle
\normalsize

\section{\label{sec:Intro} Introduction}

Quantum sensing is the procedure of measuring an unknown quantity of an observable using a quantum object as a probe. Quantum objects \textemdash\ those in which quantum-mechanical effects can manifest and be observed \textemdash\ are known to be highly sensitive to even tiny changes in the environment to which they are inevitably coupled. These changes can be so small that it is extremely challenging, or even impossible, to detect them by employing classical measurements. Consequently, the fact that the probe/sensor is quantum endows it with extreme sensitivity.  Back-action imposes a random change of a system state during the measurement. The probability of an outcome depends on the initial state of the system and on the strength of the measurement~\cite{Back-action_general}. In the case of a quantum sensor, the back-action is quantum-limited, and measurement schemes where it can be evaded have been demonstrated, e.g. in Ref~\cite{Back-action_evading}. Quantum sensors are highly engineered systems used to measure phenomena ranging from gravitational force, to magnetic and electric fields, to propagating photons. In the following, we will give a short overview of the different quantum systems that have been employed for sensing to date.

Thermal vapours of alkali atoms closed in a cell, pumped, and interrogated by near-resonant light are used to measure magnetic fields~\cite{Budker_OpticalMagnetometry}. This method is also known as nonlinear magneto-optical rotation magnetometry. Magnetometers of this type do not have intrinsic $1/f$-noise due to the absence of nearly degenerate energy states and do not require cryogenic cooling for operation; they offer millimetre spatial resolution and sensitivity exceeding ${\rm fT}/\sqrt{\rm Hz}\ $ \cite{Kominis_subfemtotelsa}. Their accuracy is shot-noise limited and scales as~\cite{Budker_magnetometry_review,Allred_SERF_magnetometry} $\delta B\sim1/\sqrt{NT_{2}t}$, where $N$ is the number of atoms, $T_{2}$ is the transverse relaxation (dephasing) time, and $t$ is the time of the signal acquisition. Spin-exchange relaxation free (SERF) operation can be achieved by increasing the gas density and improves the sensitivity of atomic magnetometers~\cite{Allred_SERF_magnetometry}. Another type of magnetic field sensor utilises ensembles of nuclear spins~\cite{Waters_magnetometer_NMR}. Although they are not as sensitive as atomic vapour sensors, they find applications in a variety of areas from archaeology to MRI systems~\cite{Degen_sensor_review} due to their simplicity and robustness. 

Nitrogen vacancy centres (NV-centres) in diamond \--- electron spin defects \--- have recently attracted a lot of attention as quantum sensors, with predicted sensitivity for ensembles of spins $\sim0.25\cdot {\rm fT}/\sqrt{{\rm Hz}\cdot {\rm cm}^3}$~\cite{Taylor_diamond_magnetometer}, and experimentally achieved sensitivities of $\sim 1 {\rm pT}/\sqrt{\rm Hz}$~\cite{Wolf_NVensemble}. With the advent of single spin in diamond readout~\cite{Gruber_single_defect_spectroscopy,Dobrovitski_single_spin_control}, it became possible to use such single spins for magnetometry~\cite{Taylor_diamond_magnetometer,Balasubramanian_nanoscale_imaging_magnetometry,Cole_decoherence_microscopy}, sensing of electric fields~\cite{Dolde_electric_field_sensing_NV_spin}, or to measure pressure~\cite{Doherty_NV_centre_pressure_sensor}. Frequency standards~\cite{Hodges_NV_centre_frequency_standard} and nanoscale thermometry with $5 {\rm mK}/\sqrt{\rm Hz}$ sensitivity based on NV defect centres in diamond have been demonstrated~\cite{Kucsko_NV_centre_thermomtry,Neumann_NV_centre_thermomtry,Toyli_NV_centre_thermometry}. The main advantages of this type of sensor are their stability in nanostructures and excellent $10-100$ nm spatial resolution.

Trapped ions have been employed to detect extremely small forces and displacements. To increase the solid angle of the field access to the trapped ion, an enhanced access ion trap geometry was shown~\cite{Maiwald_enhanced_access_ion_trap}. A force sensitivity of $\sim 100 {\rm yN}/\sqrt{\rm Hz}$ has been reached in detectors based on crystals of trapped atomic ions, which have the ability to discriminate ion displacements of $\sim 18 {\rm nm}$~\cite{Biercuk_trapped_ions_force_detection}. Their augmented force and displacement sensitivity are often traded off against the reduced resolution. Rydberg atoms are another physical system for quantum sensing of electrical fields. Their high sensitivity is based on huge dipole moments of highly excited electronic states~\cite{Osterwalder_Rydberg_states_sensing}. Rubidium atoms prepared in circular Rydberg states have been used for non-destructive (quantum nondemolition~\cite{Braginsky_QND_measurements,Levenson_QND_optics,Braginsky_QND,Grangier_QND_meaurement_optics}) measurement of single microwave photons~\cite{Nogues_QND_photon_measurement,Gleyzes_quantum_jumps_of_light}, and sensitivities reaching $3\cdot 10^{-3}\ {\rm V}/{\rm m}\sqrt{\rm Hz}$ have been achieved when Schr{\"o}dinger-cat states~\cite{Hacker_NaturePhotonics_cat_states_Rb,Vlastakis_100Photon_cat_state} were involved in the protocol~\cite{Facon_SchrodingerCat_electrometer}. The reader is directed to reviews on atomic spectroscopy- and interferometry-based sensors~\cite{Kitching_review}, quantum metrology with single spins in diamond~\cite{Chen_spins_in_diamond_review}, comparative analysis of magnetic field sensors~\cite{Lenz_magnetic_sensors_review}, and a more general and comprehensive review on quantum sensing~\cite{Degen_sensor_review}.

The article is structured as follows: Section II discusses sensors employing superconducting qubits; decoherence rates for a given design of a tunable qubit are estimated in Section III as functions of external flux and maximal qubit transition frequency; the optimal flux bias point for sensing is found in Section IV for different maximal qubit transition frequencies; Section V explores in simulation the use of entanglement between multiple qubits and a Kitaev phase estimation algorithm for sensing; finally, Section VI summarises the conclusions of this article.
 
\section{\label{sec:principle} Quantum sensors based on qubit/qudit containing circuits}

Superconducting circuits including macroscopic, human-designed, many-level anharmonic systems, operated as qubits or qudits, are a well-established experimental technology platform in the field of quantum computation and simulation. New applications of superconducting circuits comprising qubits/qudits in quantum sensing and metrology are emerging now, and the first experimental works where such circuits are used as quantum sensors have recently appeared. The frequency and amplitude of a microwave signal have been determined by spectroscopic means~\cite{Schneider_spectroscopy} and with time-domain measurements~\cite{Kristen_time_domain} via ac Stark shifts of higher qudit energy levels. The use of a persistent current qubit for ultra-high sensitivity detection of alternating current (AC) magnetic fields has been demonstrated~\cite{AC_magnetometry}. Furthermore, the absolute power flowing along a transmission line~\cite{Honigl_power_sensor} and distortions of microwave control pulses~\cite{Bylander_distortions,Gustavsson_distortions} have been measured by strong coupling to a flux qubit. Methods of using a transmon qubit as a vector network analyzer for in situ characterization of the transfer function of xy-control lines~\cite{Jerger_VNA}, and as a cryoscope to compensate for the distortions of z-control pulses~\cite{Rol_cryoscope} have been demonstrated recently. These methods are useful for the calibration of microwave lines and the measurement of the power reaching the circuit at millikelvin temperatures. They allow for the correction of pulse imperfections and increase the fidelities of control gates used in quantum computation and simulation. All of these methods are implemented on a superconducting structure comprising a single qubit/qudit.

In quantum information processing, precise dynamic control of the quantum states is key to increasing circuit depth. Conventional qubit frequency tuning is achieved by applying a well-controlled magnetic flux through a split junction loop within the quantum circuit (see Fig.~\ref{FigS1_v1}). In turn, the quantum circuit can sense these externally generated static or dynamic fields. In this article, we are using magnetic flux as an external parameter here, but the sensed quantity could also be a voltage. Electric fields are sensed by replacing the flux-threaded split junction with a voltage-biased junction (gatemon~\cite{Larsen_gatemon,Lange_gatemon}). 

Superconducting quantum circuits possess all the properties required to construct external field sensing quantum systems~\cite{Degen_review}: they have quantized energy levels; it is possible to initialize, coherently control, and readout their quantum states; and energy levels of the circuit, $E_i(\lambda)$, can be made dependent on an external parameter, $\lambda$, to be measured. For frequency-tunable qubits with a split junction, the parameter $\lambda$ is an external flux, $\Phi_{\rm{ext}}$. If the qubit is prepared in a superposition of basis states $\{0,1\}$ and placed in an external field, its state will accumulate phase $\phi(\Phi_{\rm{ext}})=\Delta\omega(\Phi_{\rm{ext}})\cdot\tau$, dependent on the flux $\Phi_{\rm{ext}}$. $\Delta\omega(\Phi_{\rm{ext}})=\omega_q(\Phi_{\rm{ext}})-\omega_d$ is the detuning between the qubit and the control pulse frequency used for the state preparation. By applying a second control pulse identical to the first one after some time $\tau$, and measuring the population of qubit basis states, it is possible to reveal the accumulated phase in oscillating dependencies of $P_{\vert 0\rangle}$ and $P_{\vert 1\rangle}$. This measurement, known as Ramsey fringes interferometry, can be employed for field sensing tasks. An equal superposition state $(\vert 0\rangle+\vert 1\rangle)/\sqrt{2}$ provides the maximal pattern visibility here, and the best sensitivity to the field. 
\begin{figure}[h]
\includegraphics[width=\columnwidth]{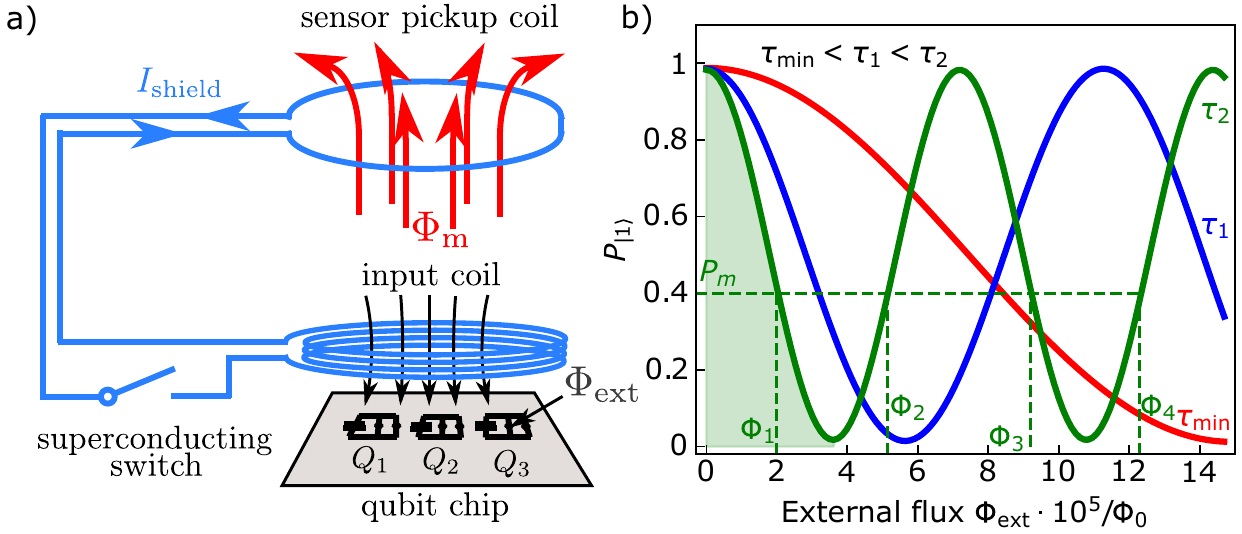}
\caption{a) Schematic diagram of a setup that can be used to employ superconducting multi-qubit structures for sensing. b) Ramsey fringes pattern $P_{\vert 1\rangle}(\Phi_{\rm{ext}},\tau)$ at different delay times $\tau_i$. $P_m$ is an outcome measured during the sensing procedure and used for the determination of the unknown external flux $\Phi$.}
\label{Fig1_v1}
\end{figure}

In practice, the external flux $\Phi_{\rm{m}}$ can be sensed by the sensor pickup coil in an input circuit converting the flux $\Phi_{\rm{m}}$ into magnetic flux $\Phi_{\rm{ext}}$ which is sensed by the qubits (Fig.~\ref{Fig1_v1}a). A fast superconducting switch \cite{NbN_switch} can be used in the input circuit to allow for the application of the flux $\Phi_{\rm{ext}}$ to the qubits only during the desired time $\tau$ with a quick on-and-off switching. After closing a superconducting switch with the $\Phi_{\rm{m}}$ flux applied, a shielding current $I_{\rm{shield}}$ will appear in the input circuit containing the pickup coil and the input coil. For small fluxes, this current will compensate $\Phi_{\rm{m}}$ to the quantized value $n\Phi_0$ with $n=0$. In this case, $I_{\rm{shield}}\sim\Phi_{\rm{m}}$ and $\Phi_{\rm{ext}}\sim\Phi_{\rm{m}}$. Similar to sensors based on dc-SQUIDs (Superconducting Quantum Interference Devices)~\cite{SQUIDs}, a variety of different input circuits can be coupled to the input coil depending on whether a voltmeter, magnetometer, or gradiometer needs to be produced. The pick-up coil, sketched in Fig.~\ref{Fig1_v1}a, has an inductance, which screens the high frequency magnetic field fluctuations from the quantum chip. Only lower frequencies will pass and been seen by the qubit arrays, the cut-off frequencies are set by the circuit design. Measurement sequences within the inverse cut-off frequency will see a stable (averaged) magnetic field, and the measurements are affected by low frequency magnetic field noise indeed. In the following, we focus on sensing of the flux $\Phi_{\rm{ext}}$ applied to the qubits. 

The Ramsey fringes pattern $P_{\vert 1\rangle}(\Phi_{\rm{ext}},\tau)$ can be simulated or directly measured as a calibration pattern before the field sensing routine. In this scenario, the outcome $P_m$ measured during the sensing procedure will be used in conjunction with the calibration pattern to determine the unknown flux value. Fig.~\ref{Fig1_v1}b shows the simulated dependence of the probability $P_{\vert 1\rangle}(\Phi_{\rm{ext}},\tau)$ on the external flux $\Phi$ at different delay times $\tau_i$. One can see that the longer the delay time, the higher the sensitivity of $P_{\vert 1\rangle}$ to the external flux. This is only the case if the delay time $\tau$  is shorter than the coherence time, $T_2$, of the qubit; for longer delay times, the sensitivity will be reduced. Two issues should be noted here. Firstly, for an unknown flux value, it is not possible to choose a priori the delay time $\tau_{\rm{opt}}$ with the best sensitivity. Secondly, for longer delay times, it is not possible to unambiguously determine the measured flux based on a single outcome. As shown in Fig.~\ref{Fig1_v1}b, the same result $P_m$ can correspond to many flux values $\{\Phi_1,\Phi_2,\Phi_3,\Phi_4\}$. To make the measurement unambiguous, one has to reduce the dynamic range of the sensor to the interval highlighted in green. This interval is substantially shorter than for the shortest delay time $\tau_{\rm{min}}$, where the measurement is single-valued (one-to-one correspondence).

Phase estimation algorithms~\cite{Giovannetti_2011,Giovannetti_2006} (PEA)  address both issues. They gradually tune the delay time to the value $\tau_{\rm{opt}}\sim T_2$ with the highest available sensitivity, without a reduction in the dynamic range of the sensor, and appear to be powerful tools in sensing. Kitaev~\cite{Kitaev_1995} and Fourier~\cite{van_Dam_2007} phase estimation algorithms have been used with a single tunable transmon qubit to measure external flux, and the scaling of accuracy beyond the standard quantum limit (SQL) has been experimentally demonstrated~\cite{Danilin_magnetometry}. These algorithms involve a stepped strategy. The Kitaev algorithm starts at the minimal delay time $\tau_{\rm{min}}$. At each step of the algorithm, the interval of possible fluxes is reduced by a factor of two based on the measurement outcome, and a new optimal delay time is found for the next step providing improved sensitivity. The delay times grow from step to step on average and gradually tend to the optimal delay time $\tau_{\rm{opt}}\sim T_2$. PEAs allow us to approach $\sim 1/t$ accuracy scaling \--- the Heisenberg limit (HL) where $t$ is the sensing limit. The optimal delay time $\tau_{\rm{opt}}$ is of the order of the coherence time, see Sec.~\ref{sec:optimization}, and the qubit coherence time $T_2$ serves as a quantum resource. The longer it is, the more steps of the algorithm that can be completed before the delay time approaches $\tau_{\rm{opt}}$ and a higher accuracy can be achieved in the same sensing time. Improvement strategies exploit classical machine learning algorithms to improve the precision of quantum phase estimation~\cite{machine_learning_Paraoanu} and   quantum sensing algorithms employing qutrits instead of qubits~\cite{Shlyakhov_2018} have also been considered.

\section{\label{sec:Sec2} Coherence of the sensor}

To determine the optimal sensing point, we estimate the relaxation and dephasing rates. While the analysis is general, we provide specific data for one of the most common qubit designs shown in Fig.~\ref{FigS1_v1}. We perform numerical simulations to compute the parameters required for the estimation of the qubit decoherence rates (see ~\ref{A1} for the details). The qubit $0\--1$ transition frequency can be parameterised by its maximal value $f_q^{\rm{max}}$ and external magnetic flux $\Phi$ as
\begin{equation}
f_q(f_q^{\rm{max}},\Phi) = \left(f_q^{\rm{max}}+\frac{E_C}{h}\right)\sqrt{\vert\cos\left(\pi\frac{\Phi}{\Phi_0}\right)\vert}-\frac{E_C}{h},
\label{eq_1}
\end{equation}
where $\Phi_0$ is the magnetic flux quantum. We limit the variation of the external flux to the first half-period $\Phi/\Phi_0\in [0,0.5]$. In this case, the absolute value of the cosine under the square root can be omitted. The maximal value of the transition frequency $f_q^{\rm{max}}$ depends on the charging energy $E_C$ and on the maximal Josephson energy $E_J$. This allows us to treat it as a variable here, different values of which can be targeted during the fabrication of the device by aiming for different critical currents of the Josephson junctions.

Next, we analyse the coherence limitation for an x-mon type superconducting qubit. 
\subsection{\label{subsec:relaxation} Energy relaxation}

\subsubsection{Purcell relaxation.} The Purcell relaxation rate of the qubit depends on the qubit-resonator detuning $\Delta = \vert\omega_q-\omega_r\vert$. If the multi-mode model for the resonator is employed~\cite{Purcell_rate}, it is not symmetric with respect to the readout resonator frequency $\omega_r$. For simplicity, we consider the qubit sensor operated at a frequency below the resonator where the relaxation rates are substantially lower. To keep the fidelity of the readout the same for all possible points of operation, a qubit-resonator detuning $\Delta=2\pi\times 2\ \rm{GHz}$ is kept constant. The resonator single-mode formula for the relaxation rate in the resonator reads
\begin{equation}
    \Gamma_1^{\rm{cav}}(f_q^{\rm{max}},\Phi)=\kappa\frac{g_{01}^2(f_q^{\rm{max}},\Phi)}{\Delta^2}.
    \label{Purcell_rate}
\end{equation}

Eq.~(\ref{Purcell_rate}) gives us an overestimated relaxation rate~\cite{Purcell_rate} and provides a conservative estimate. We will use a total resonator decay rate of $\kappa=0.5\ \rm{MHz}$ for our computations as this is a good representation of readout resonator width. The qubit-resonator coupling strength $g_{01}$ in Eq.~(\ref{Purcell_rate}) depends on the maximal qubit frequency $f_q^{\rm{max}}$ and external flux $\Phi$ as
\begin{equation}
  g_{01} = \beta e\sqrt{\frac{Z_0}{h}}(\omega_q+\Delta)\sqrt{\frac{hf_q^{\rm{max}}+E_C}{E_C}}\cos^{\frac{1}{4}}\left(\pi\frac{\Phi}{\Phi_0}\right),
\label{eq_3}
\end{equation}
where $Z_0 = 50\ \Omega$ is the characteristic impedance of the resonator.

\subsubsection{Losses in dielectrics and due to quasiparticle tunneling.} The energy dissipation in dielectric materials does not depend either on the frequency in the $1-10\ \rm{GHz}$ range or on the external magnetic flux. With proper choice of materials and circuit designs this relaxation channel can be reduced dramatically~\cite{Losses_res}. The relaxation rate due to quasiparticle tunneling grows as $\sim\omega_q^{-1/2}$ when the flux approaches $\Phi_0/2$ and the qubit transition frequency $\omega_q$ is lowered~\cite{Koch_2007}, but its value is in the range of a few $\rm{Hz}$ at $\sim\ 20\ \rm{mK}$ temperature~\cite{Koch_2007} and does not contribute noticeably to the total relaxation.

\subsubsection{Spontaneous emission.} In contrast to the traditional transmon geometry with two floating electrodes, for an x-mon design, the ground plane completely surrounds the qubit electrode. During the oscillations of charge related to different qubit states, the centre of the positive charges will always coincide with the centre of the negative charges. As a result, the dipole moment of the x-mon qubit is zero in the first approximation, and the relaxation decay channel related to the radiation losses can be neglected: $\Gamma_1^{\rm{rad}}=0$. 

\subsubsection{Relaxation due to the inductive coupling to the z-control line.} We perform an estimation of the mutual inductance $M=2.08\ \rm{pH}$ between the flux bias line and the x-mon SQUID loop of inductance, and between the flux bias line and the total circuit of the x-mon $M^\prime$, see Fig.~\ref{FigS1_v1}. This inductance $M^\prime=0.22\ \rm{pH}$ is related to the total flux threading the gap between the x-mon electrode and the ground plane. It arises from the asymmetric placement of the z-control line with respect to the arm of the x-mon (see more detail in ~\ref{A1}). The relaxation rate related to the inductive coupling to the z-control line can be computed as 
\begin{equation}
\Gamma_1^{\rm{ind}}(f_q^{\rm{max}},\Phi)=\frac{(M^2+M^{\prime 2})\omega_q^2(f_q^{\rm{max}},\Phi)}{L_J(f_q^{\rm{max}},\Phi)Z_0}.
\label{inductive_coupling_rate}
\end{equation}
Here, $L_J=\hbar^2/(4e^2E_J)$ is the Josephson inductance and $Z_0=50\ \Omega$ is the impedance of the z-control line. The Josephson inductance depends on the maximal qubit transition frequency and the external flux as
\begin{equation}
L_J(f_q^{\rm{max}},\Phi)=\frac{2E_C}{e^2(\omega_q^{\rm{max}}+\frac{E_C}{\hbar})^2\cos\left(\pi\frac{\Phi}{\Phi_0}\right)}.
\label{Josephson_inductance}
\end{equation}

\subsubsection{Relaxation due to the capacitive coupling to the xy-control line.} The xy-control line can also be called the gate line. It is used to apply fast microwave pulses to the qubit at a specific transition frequency $f_q(\Phi)$ to change its quantum state. Due to the capacitive coupling between this line and the qubit, the energy is dissipated through this channel, and the relaxation rate can be estimated as
\begin{equation}
\Gamma_1^{\rm{cap}}(f_q^{\rm{max}},\Phi)=\frac{\omega_q^2(f_q^{\rm{max}},\Phi)Z_0C_c^2}{C_{qg}}.
\label{capacitive_coupling_rate}
\end{equation}
The coupling capacitance, $C_c=0.2\ \rm{fF}$, between the x-mon electrode and the xy-control line and the capacitance, $C_{qg}=76\ \rm{fF}$, between the x-mon electrode and the ground plane were computed numerically for the geometry given in Fig.~\ref{FigS1_v1} (see ~\ref{A1} for the details).

\subsection{\label{subsec:dephasing} Pure dephasing}

\subsubsection{Charge noise and quasiparticle tunneling.} The exponential decrease in the charge dispersion of the transmon device with $\sqrt{E_J/E_C}$~\cite{Koch_2007} sets the energy spectrum of the device internally immune to the {\it charge noise} and {\it quasiparticle tunneling}. The dephasing caused by these mechanisms can be completely neglected.   

\subsubsection{Flux noise.} This type of noise plays a crucial role in the optimization of sensor operation as the external flux measured by the sensor couples to the device in the same way as the flux noise. The low-frequency part of the noise spectrum with frequencies lower than the qubit transition frequency has the most significant contribution to the dephasing of the qubit. Recent studies of decoherence in superconducting qubits~\cite{Decoherence_Burnett,Decoherence_Schlor} suggest that the noise spectrum is of $1/f$-type at low frequencies, so we will use the noise power spectral density $S_\Phi(\omega)=2\pi A^2_\Phi/\omega$ with $A_\Phi=10^{-6}\Phi_0=\alpha\Phi_0$~\cite{Wellstood_flux, Yoshihara_flux} for our estimations. The linear and quadratic flux noise contributions to the Ramsey fringes decay function translate to the noise in the qubit transition frequency $\delta\omega_q \simeq \vert\partial\omega_q/\partial\Phi\vert\delta\Phi+\vert\partial^2\omega_q/\partial\Phi^2\vert\delta\Phi^2/2$. This leads to the factor-function~\cite{Koch_2007,Decoherence_Ithier} 
\begin{equation}
\rm{exp}\left(-t^2A^2_\Phi\left|\frac{\partial\omega_q}{\partial\Phi}\right|^2\right)\rm{exp}\left(-t\pi^2A^2_\Phi\left|\frac{\partial^2\omega_q}{\partial\Phi^2}\right|\right)
\label{decay_factor_flux}
\end{equation} in the decay.
The linear contribution results in a Gaussian decay factor-function, and the quadratic contribution leads to the exponential decay. Expressed through the parameters of Eq.~(\ref{eq_1}) this factor-function reads
\begin{equation}
\begin{aligned}
\rm{exp}\left(-\frac{t^2\alpha^2\pi^4(f_q^{\rm{max}}+\frac{E_C}{h})^2\sin^2\left(\pi\frac{\Phi}{\Phi_0}\right)}{\cos\left(\pi\frac{\Phi}{\Phi_0}\right)}\right)\times\\
\rm{exp}\left(-\frac{t\pi^5\alpha^2(f_q^{\rm{max}}+\frac{E_C}{h})(1+\cos^2\left(\pi\frac{\Phi}{\Phi_0}\right))}{2\cos^{\frac{3}{2}}\left(\pi\frac{\Phi}{\Phi_0}\right)}\right).
\end{aligned}
\label{eq_8}
\end{equation}

\subsubsection{Critical current noise.} Fluctuations in the critical current of the junctions $\delta I_c$ lead to the noise in the maximal Josephson energy of the qubit and in its transition frequency $\delta\omega_q\simeq\vert\partial\omega_q/\partial I_c\vert\delta I_c$. In terms of Eq.~(\ref{eq_1}), we can express the derivative as
\begin{equation}
\frac{\partial\omega_q}{\partial I_c}=\frac{4E_C}{(E_C+hf^{\rm{max}}_q)e}\sqrt{\cos\left(\pi\frac{\Phi}{\Phi_0}\right)}.
\label{eq_9}
\end{equation}
Noise in the critical current is also of 1/f-type at low frequencies~\cite{crit_curr_noise} with noise power spectral density $S_{I_c}(\omega)=2\pi A^2_{I_c}/\omega$, $A_{I_c}=10^{-6}I_c=\gamma I_c$ ~\cite{crit_curr_noise}. Dephasing due to this critical current noise results in a Gaussian decay factor-function $\rm{exp}\{-t^2A^2_{I_c}\left(\partial\omega_q/\partial I_c\right)^2\}$ in the Ramsey fringes experiment, which can be expressed as
\begin{equation}
\rm{exp}\left(-t^2\pi^2\gamma^2\left(f^{\rm{max}}_q+\frac{E_C}{h}\right)^2\cos\left(\pi\frac{\Phi}{\Phi_0}\right)\right).
\label{eq_10}
\end{equation}

\begin{table*}[ht]
\centering
  \begin{tabular}{|p{1.7cm}|p{1.7cm}|p{1.7cm}|p{1.7cm}|p{1.7cm}|p{1.7cm}|p{1.7cm}|}
  \hline
    $\Phi/\Phi_0$ & $\Gamma_1^{\rm{cav}}$& $\Gamma_1^{\rm{ind}}$& $\Gamma_1^{\rm{cap}}$& $\Gamma_{\phi,1}^{\rm{flux}}$& $\Gamma_{\phi,2}^{\rm{flux}}$& $\Gamma_\phi^{\rm{curr}}$\\
    \hline
    0 & 5.0 & 71.9 & 99.3 & 2.8$\cdot 10^{-3}$ & 0 & 29.1 \\
    \hline
    0.2 & 3.4 & 46.8 & 79.8 & 3.2$\cdot 10^{-3}$ & 59.7 & 26.1 \\
    \hline
    0.4 & 0.6 & 6.6 & 29.3 & 9.0$\cdot 10^{-3}$ & 156.3 & 16.2 \\
    \hline
  \end{tabular}
 \caption{Estimates of the relaxation and pure dephasing rates in kHz for the considered transmon qubit design (see ~\ref{A1}) for 3 different flux values. The pure dephasing rates arising from the flux noise are the only decoherence rates which grow when the flux is detuned from the ``sweet spot" $\Phi=0$.}
\label{tab_1}
\end{table*}

\section{\label{sec:optimization} Optimal sensing point}

During the sensing procedure we measure the outcome $P_m$, the probability of finding the qubit in the first excited state, at a known delay time $\tau$ between the control pulses and convert it to the magnetic flux $\Phi_{\rm{ext}}$ by using the $P_{|1\rangle}(\Phi_{\rm{ext}},\tau)$ pattern. The more accurately we measure the outcome $P_m$, the better we know the flux $\Phi_{\rm{ext}}$, $\delta P_m=\partial P_{\vert 1\rangle}/\partial\Phi\cdot\delta\Phi$. To determine the outcome $P_m$, we have to prepare and measure the qubit state some $K$ times, and the accuracy of the outcome will drop down as $\delta P_m\le 1/(2\sqrt{K})$ \cite{Danilin_magnetometry}. The accuracy of the flux then scales as $\delta\Phi\le 1/[2\sqrt{K}(\partial P_{\vert 1\rangle}/\partial\Phi)]$. Every iteration of state preparation and measurement takes some time $T_r$, set by the experimental setup, and the total time of sensing $t=K\cdot T_r$. As a result, the flux accuracy can be expressed as
\begin{equation}
\delta\Phi\le\frac{1}{2\sqrt{\frac{t}{T_r}}\left(\frac{\partial P_{\vert 1\rangle}}{\partial\Phi}\right)}=\frac{A}{\sqrt{t}},
    \label{sensitivity_eq}
\end{equation}
where $A$ represents the sensitivity of the sensor and has units of $[\Phi_0/\sqrt{\rm{Hz}}]$.

For optimal sensing of external fields we need to determine the operational point in flux $\Phi^*$ where the Ramsey fringe pattern is the most sensitive to any changes in the external flux $\delta\Phi$. Minimal $A$ is reached when the $\partial P_{\vert 1\rangle}/\partial\Phi$ is maximal, this can be expressed as 
\begin{equation}
\max_{(\Phi,\tau)}\left(\frac{\partial P_{\vert 1\rangle}}{\partial\Phi}\right)_{\rm{amp}}.
\label{eq_11}
\end{equation}
We describe the Ramsey fringe pattern of the qubit excited state probability as
\begin{equation}
P_{\vert 1\rangle}(\Phi,\tau)=\frac{1}{2}+\frac{1}{2}e^{-A\tau-B^2\tau^2}\cos(\Delta\omega(\Phi) \tau),
\label{eq_12}
\end{equation}

\begin{figure*}[t]
\includegraphics[width=\textwidth]{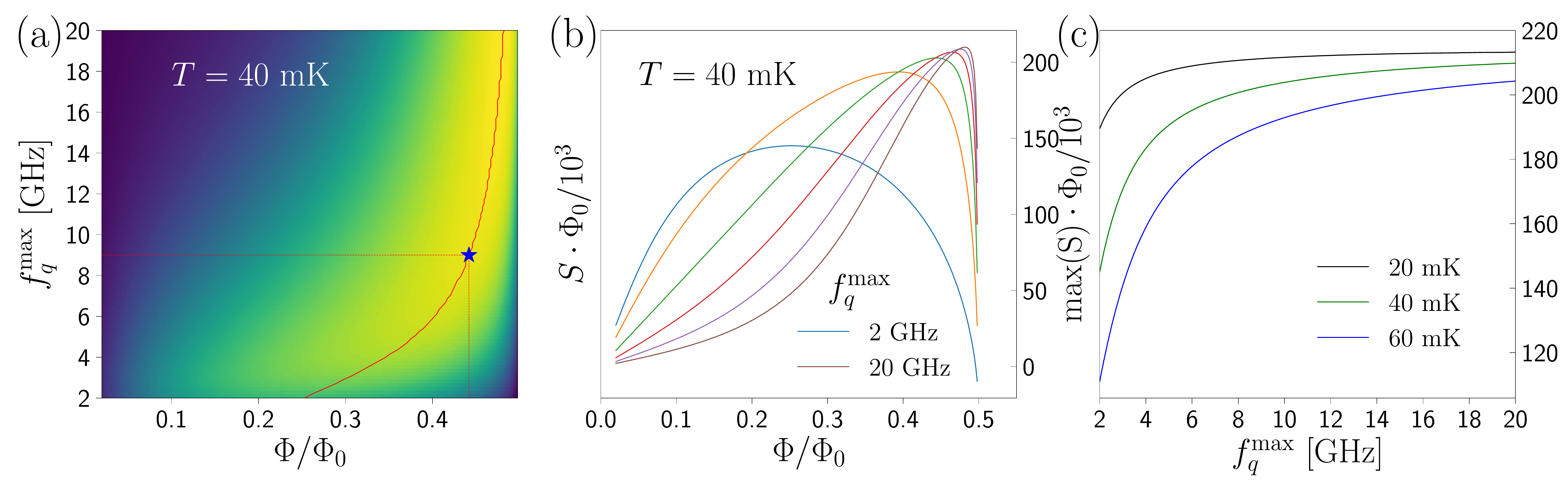}
\caption{Most sensitive working points. (a) Sensitivity of the $P_{\vert 1\rangle}(\Phi,\tau)$ pattern at the optimal delay time $\tau_{\rm{opt}}$ to the changes of external flux $\Phi$ at the temperature of $40\ \rm {mK}$ as a function of $\Phi/\Phi_0$ for sensors with different maximal $0 - 1$ transition frequencies $f_q^{\rm{max}}$. Red line shows the optimal flux for every frequency $f_q^{\rm{max}}$. Blue star denotes the optimal for sensing flux for a sensor with $f_q^{\rm{max}}=9\ \rm{GHz}$. (b) The same sensitivity as in (a) shown as a function of $\Phi/\Phi_0$ for 6 different $f_q^{\rm{max}}$ frequencies equally spaced between the $2\ \rm{GHz}$ and $20\ \rm{GHz}$. The maximal sensitivity grows with the $f_q^{\rm{max}}$ frequency. (c) Sensitivities reached along the ridge of the function in (a) as a function of $f_q^{\rm{max}}$ frequency. For $T=40\ \rm{mK}$, this corresponds to $S\cdot\Phi_0$ values along the red line in (a).}
\label{Fig2}
\end{figure*}
where $A$ and $B$ are the decoherence rates of the sensor, $\Delta\omega=\omega_q(\Phi)-\omega_d$ is the detuning between the qubit and the drive pulses frequencies, and $\tau$ is the delay time between the control pulses in the Ramsey fringes experiment. The decoherence rates depend on the rates defined earlier in Eq.~(\ref{Purcell_rate},\ref{inductive_coupling_rate},\ref{capacitive_coupling_rate},\ref{eq_8},\ref{eq_10}),
\begin{equation}
\begin{aligned}
A = \frac{\Gamma_1^{\rm{cav}}+\Gamma_1^{\rm{ind}}+\Gamma_1^{\rm{cap}}}{2}+\pi^2 A_\Phi^2\left|\frac{\partial^2\omega_q}{\partial\Phi^2}\right|,\\
B= \sqrt{A_\Phi^2\left|\frac{\partial\omega_q}{\partial\Phi}\right|^2+A_{I_c}^2\left|\frac{\partial\omega_q}{\partial I_c}\right|^2}.
\end{aligned}
\label{eq_13}
\end{equation}
The quickest change of the excited state probability in Eq.~(\ref{eq_12}) arises from the cosine term, and taking the derivative in flux $\Phi$, we keep the rates $A$ and $B$ constant and given by the values of the maximal qubit transition frequency $f_q^{\rm{max}}$ and $\Phi$. Taking into account that $\partial\Delta\omega/\partial\Phi=\partial\omega_q/\partial\Phi$, we find
\begin{equation}
\left(\frac{\partial P_{\vert 1\rangle}}{\partial\Phi}\right)_{\rm{amp}}=\frac{\tau}{2}e^{-A\tau-B^2\tau^2}\left|\frac{\partial\omega_q}{\partial\Phi}\right|.
\label{eq_14}
\end{equation}
The derivative $\partial P_{|1\rangle}/\partial\Phi$ has a sine term, but we are maximizing the amplitude of that sine function Eq.~(\ref{eq_14}) in the delay time $\tau$ between the control pulses for that $P_{|1\rangle}$ pattern to get the most sensitive to the flux $\Phi$. We find the optimal delay time 
\begin{equation}
\tau_{\rm{opt}}=\frac{-A+\sqrt{A^2+8B^2}}{4B^2}.
\label{eq_15}
\end{equation}
Substituting this $\tau_{\rm{opt}}$ time into Eq.~(\ref{eq_14}), we find the function $\left(\partial P_{\vert 1\rangle}/\partial\Phi\right)_{\rm{amp}}(f_q^{\rm{max}},\Phi)$ and search for the values of $f_q^{\rm{max},*}$ and $\Phi^*$ which maximize it. Importantly, $\Gamma_1^{\rm{cav}}, \Gamma_1^{\rm{ind}}$, and $\Gamma_1^{\rm{cap}}$ all drop down when the sensor is detuned from the ``sweet spot" at $\Phi=0$ (see Table~\ref{tab_1}), the same applies to the pure dephasing caused by critical current noise. The pure dephasing related to the flux noise, Eq.~(\ref{decay_factor_flux}), is the only decoherence rate which grows when the flux is detuned from the ``sweet spot" and tends to $\Phi=\Phi_0/2$, with the major contribution coming from the part of the exponent argument that is quadratic in time. The sensitivity of the $P_{\vert 1\rangle}$ pattern to the flux, Eq.~(\ref{eq_14}), depends linearly on the slope of the qubit spectrum in flux $\partial\omega_q/\partial\Phi$, which grows to infinity when $\Phi\rightarrow\Phi_0/2$. This growth prevails over the reduction in coherence caused by the flux noise, and the sensitivity of the sensor will also grow as the flux approaches $\Phi_0/2$. 

The qubit transition frequency $f_q$ drops down rapidly when $\Phi\rightarrow\Phi_0/2$, and thermal energy $k_BT$ becomes comparable with the transition energy of the qubit $hf_q$. In this case, we have to consider the qubit's thermal excitation which will populate the first excited state. This excitation will reduce the contrast of the Ramsey fringe pattern and lower the maximal sensitivity to the flux of the $P_{\vert 1\rangle}$ pattern. This effect can be accounted for by introducing an additional factor to the Ramsey fringe pattern, which modifies Eq.~(\ref{eq_12}) to
\begin{equation}
P_{\vert 1\rangle}=\frac{1}{2}+\frac{1}{2}\frac{e^{\frac{hf_q}{k_BT}}-1}{e^{\frac{hf_q}{k_BT}}+1}e^{-A\tau-B^2\tau^2}\cos(\Delta\omega \tau),
\label{eq_16}
\end{equation}
where $k_B$ is Boltzmann constant, and $T$ is the temperature of the sensor.
Taking the derivative of Eq.~(\ref{eq_16}) in flux, we simplify with an additional pre-factor being a constant as the change of the qubit frequency which causes the cosine argument to change by $\pi$ changes the $hf_q/k_BT$ ratio by only $0.4\%$ when qubit frequency is $f_q=1\ \rm{GHz}$. This means that the optimal time delivering the maximal sensitivity to the flux stays the same, and the maximal sensitivity value reads
\begin{equation}
\begin{aligned}
S(f_q^{\rm{max}},\Phi)=\max_{\tau}\left(\frac{\partial P_{\vert 1\rangle}}{\partial\Phi}\right)_{\rm{amp}}=\\\frac{\tau_{\rm{opt}}}{2}\frac{e^{\frac{hf_q}{k_BT}}-1}{e^{\frac{hf_q}{k_BT}}+1}e^{-A\tau_{\rm{opt}}-B^2\tau_{\rm{opt}}^2}\left|\frac{\partial\omega_q}{\partial\Phi}\right|.
\label{eq_17}
\end{aligned}
\end{equation}
This sensitivity multiplied by the flux quantum $\Phi_0$ is plotted in Fig.~\ref{Fig2}(a) for the temperature $T=40\ \rm{mK}$ as a function of maximal qubit transition frequency $f_q^{\rm{max}}$ and normalized flux $\Phi/\Phi_0$. For every value of $f_q^{\rm{max}}$, an optimal sensing point exists in flux where the sensitivity is maximal. The higher the $f_q^{\rm{max}}$ value, the closer the optimal flux point to $\Phi_0/2$ (see Fig.~\ref{Fig2}(b)). The range of $f_q^{\rm{max}}$ frequencies studied here can be accessed by the microwave generators usually used in experiments with superconducting circuits. Fig.~\ref{Fig2}(c) shows the maximally achievable sensitivities reached along the ridge of the $S(f_q^{\rm{max}},\Phi)$ function for different $f_q^{\rm{max}}$ values for three different temperatures. For $T=40\ \rm{mK}$, this corresponds to the function values reached along the red line in Fig.~\ref{Fig2}(a). The maximally achievable sensitivity does not grow much after the $f_q^{\rm{max}}$ frequency is increased above $10\ \rm{GHz}$. It is important to thermally anchor the sensor properly, as a lower temperature yields a higher sensitivity at the same $f_q^{\rm{max}}$ frequency. Though the maximal qubit transition frequency $f_q^{\rm{max}}$ can be high, for sensing the qubit will be tuned to the flux point where its frequency is below $5\ \rm{GHz}$. The resonator frequency will be above the qubit by $\Delta/2\pi\simeq 2\ \rm{GHz}$, which is in the range of operation of $4$K HEMT amplifiers. 

\begin{figure}[!h]
\includegraphics[width=\columnwidth]{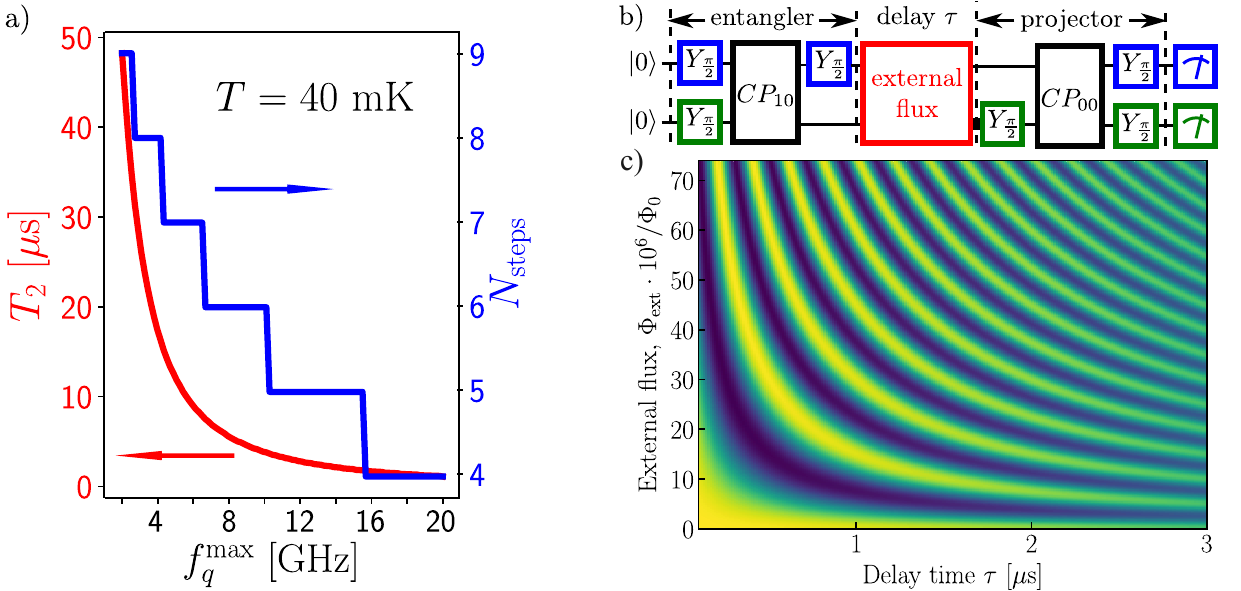}
\caption{a) Coherence time $T_2$ of the qubit sensor at the optimal sensitivity point (left axis) and the number of the Kitaev algorithm steps for a single qubit before reaching the optimal delay time $\tau_{\rm{opt}}$ (right axis) as a function of the maximal qubit transition frequency $f_q^{\rm{max}}$. b) Sequence of control operations used for sensing external fluxes with phase estimation algorithms. c) Simulated probability pattern $P_{\vert 10\rangle}(\Phi_{\rm{ext}},\tau)$ of the two-qubit state $\vert 10\rangle$ by the end of the sequence shown in b). The simulation is done for the case when both qubits have identical transition frequency spectra with $f_q^{\rm{max}}=9\ \rm{GHz}$ and for the optimal sensing point $\Phi^*/\Phi_0=0.442$ found in Sec.~(\ref{sec:optimization}) with the decoherence rates from Sec.~(\ref{sec:Sec2}).}
\label{Fig3}
\end{figure}

The coherence of the sensor at the optimal point in flux drops with the maximal frequency of the qubit sensor $f_q^{\rm{max}}$, see Fig.~\ref{Fig3} left axis. This results in reducing the maximal number of steps of the phase estimation algorithm before reaching the optimal sensing time $\tau_{\rm{opt}}$. If the minimum time delay between the control pulses is $\tau_{\rm{min}}$, the number of algorithm steps can be estimated as
\begin{equation}
N_{\rm{step}}=\lfloor\log_2\left(\frac{2\tau_{\rm{opt}}(f_q^{\rm{max}},\Phi)}{\tau_{\rm{min}}}\right)\rfloor.
\label{eq_18}
\end{equation}
Here, we assume that the delay time between the control pulses doubles as we progress from one step of the algorithm to the next one. The symbols $\lfloor$ and $\rfloor$ denote rounding of the value to the closest integer number lower than the logarithm value. The minimal delay time between the control pulses $\tau_{\rm{min}}$ is given by the speed of turning the external fields on and off. The number of the phase estimation algorithm steps for maximal qubit frequency and corresponding flux point given by the red line in Fig.~\ref{Fig2}(a) is shown in Fig.~\ref{Fig3}(a) right axis.

For a sensor with a given maximal qubit transition frequency $f_q^{\rm{max}}$, the sensitivity grows more and more as the flux is tuned away from the ``sweet spot" $\Phi=0$ until it reaches a maximum marked by the red line in Fig.~\ref{Fig2}(a). In the flux interval where the sensitivity is constantly growing, that is to the left of the red line, the dynamical range of the sensor $\Phi_{\rm{max}}$ goes down with the increase in flux $\Phi$ because of the rise in the derivative value $\vert d\omega_q/d\Phi\vert$. The dynamical range of the sensor is determined by the measurable magnitudes of the external fields. This range spans from zero to the maximal value $\Phi_{\rm{max}}$, and 
\begin{equation}
\Phi_{\rm{max}} = \frac{\pi}{\tau_{\rm{min}}\vert d\omega_q/d\Phi\vert}.
\label{eq_19}
\end{equation}
As a result, the increase in  sensitivity is balanced by the reduced range of magnitudes of the external field.

For our simulations, we choose a qubit sensor with the maximal transition frequency $f_q^{\rm{max}}=9\ \rm{GHz}$. This choice results in an optimal flux value for sensing of $\Phi^*/\Phi_0=0.442$ at $T=40\ \rm{mK}$, where the coherence time of the sensor is $T_2=4.625\ \mu\rm{s}$, the optimal delay time between the control pulses is $\tau_{\rm{opt}}=3.292\ \mu\rm{s}$, and it is possible to perform $N_{\rm{step}}=6$ steps of the Kitaev algorithm. The delay time $\tau$ between the control pulses in the algorithm is the time during which the quantum state of the sensor accumulates its phase $\phi(\Phi_{\rm{ext}})$. Throughout the stepped algorithm, the quantum sensor is exposed to the measured field many times and the total phase accumulation time can be computed. This time is not dependent on the instrumental details of the sensor and is a good metric against which the scaling of the flux accuracy can be plotted. The black dots in Fig.~\ref{Fig6}(c) show how the averaged accuracy of the flux measurement $\overline{\delta\Phi/\Phi_0}$ scales with the averaged total phase accumulation time $\bar{\tau}$ for a Kitaev PEA run on a single qubit sensor with the parameters described above.

\section{\label{ent_sens}Adding Entanglement}

Quantum entanglement can provide improvements in attainable sensitivity~\cite{Giovannetti_2004,Giovannetti_2012} for short interrogation times $\tau$ since the entangled state of $N$ qubits used as probes allows for an N-fold speed-up in phase accumulation. Experimentally, this has been demonstrated for systems including three trapped $^9$Be$^+$ ions~\cite{Leibfried_2004}, four-entangled photons~\cite{Nagata_2007}, ten nuclear spins~\cite{Jones_2009}, and a single bosonic mode of a superconducting resonator~\cite{Wang_2019}. Though these experiments clearly demonstrate the improvement of sensitivity beyond the SQL with the number
of entangled qubits, they do not yet provide an explicit metrologic routine for the measurement of an unknown external field. To this end, we suggest using the probability pattern $P_{\vert 10\rangle}$ of a two-qubit state for sensing with phase estimation algorithms (PEAs). Fig.~\ref{Fig3} (b) shows an example of a time sequence of gates which can be used for flux sensing. It employs the conditional phase two-qubit gates~\cite{cPhase_gate}, and $CP_{ij}$ denotes that conditional phase gate which inverts the sign of only the $\vert ij\rangle$ state. The flux dependence of the qubit transition frequency is assumed to be the same for both qubits with $f_q^{\rm{max}}=9\ \rm{GHz}$. They are both tuned to the flux point $\Phi^*/\Phi_0=0.442$ where they have their maximum sensitivity. Starting from both qubits in the ground state,
we create the $\vert\Phi^+\rangle$ Bell state, apply the external flux we want to measure to both qubits, and allow the system to evolve for a variable time $\tau$. After that, we convert the entangled state to a separable state, projecting the entangled state phase to the phase of the first qubit, shown in Eq.~(\ref{eq_20}).

\begin{multline}
|00\rangle\xrightarrow[CP_{10}]{\rm Entangler}\frac{|00\rangle+|11\rangle}{\sqrt{2}}\underset{\tau}{\Longrightarrow}\frac{|00\rangle+e^{i\phi(\Phi_{\rm ext},\tau)}|11\rangle}{\sqrt{2}}\\
\xrightarrow[CP_{00}]{\rm Projector}\left(\frac{-1+e^{i\phi}}{2}|0\rangle+\frac{-1-e^{i\phi}}{2}|1\rangle\right)\otimes|0\rangle.
\label{eq_20}
\end{multline}
Subsequent measurement of both qubit states, for different delay times $\tau$ and different external fluxes $\Phi_{\rm{ext}}$, allow the determination of the probabilities of all four possible two-qubit states. The pattern $P_{\vert 10\rangle}$ is shown in Fig.~\ref{Fig3}(c). It closely resembles that of the Ramsey fringes, but has double the frequency of $P_{\vert 10\rangle}$ oscillations, $\phi = 2\times\Delta\omega\times\tau$. The doubling of phase accumulation speed results in two times better accuracy of flux sensing at the same short sensing times. However, the pattern contrast also reduces quicker in comparison with a single qubit case, making the advantage less impressive for long measurement times. This originates from the shortening of the coherence time $T_2$ with the growth of the system size $N$. In our simulations, we assume that the decoherence rates grow proportionately to the number of qubits used for sensing $\Gamma_N = N\Gamma$. In general, the pure dephasing rate is proportional to $\sim N^\alpha$, with $\alpha=1$ representing non-correlated noise and $\alpha=2$ representing correlated noise acting on all qubits. Experimental investigations into noise correlations between two or more superconducting qubits have only recently started to appear~\cite{correlations_1,correlations_2,correlations_3}, being important for quantum computation and quantum-enhanced sensing.

Next, we simulate the flux sensing routines based on the Kitaev PEA run with a single qubit, and with two and three qubits prepared in the GHZ entangled state. We compare the accuracy of flux sensing achieved by employing entangled states to that of a single qubit. To perform the simulation we
compute the probability patterns $P_{\vert 10...0\rangle}$ of the $N$-qubit states for $N=1,2,\ \rm{and}\ 3$ as
\begin{equation}
P_{\vert 10..0\rangle}=\frac{1}{2}+\frac{1}{2}\frac{e^{\frac{hf_q}{k_BT}}-1}{e^{\frac{hf_q}{k_BT}}+1}e^{-N(A\tau+B^2\tau^2)}\cos(N\Delta\omega\tau).
\label{eq_21}
\end{equation}
These probabilities are obtained after projecting the phase accumulated by the GHZ $N$-qubit state during the evolution in the external magnetic field to the first qubit. The dependencies of the qubits’ spectra on the flux are assumed to be identical with the maximal transition frequency $f_q^{\rm{max}}=9\ \rm{GHz}$, and all qubits are tuned to the optimal sensing point at $\Phi^*/\Phi_0=0.442$. The total decoherence rates for $N$-qubit entangled states are considered to grow proportionately to $N$. We use the equidistant flux grids in the computation of probability patterns with 6144, 3072, and 2048 flux values for 1-, 2-, and 3-qubit cases, respectively. 
\begin{figure}[!h]
\centering
\includegraphics[width=\columnwidth]{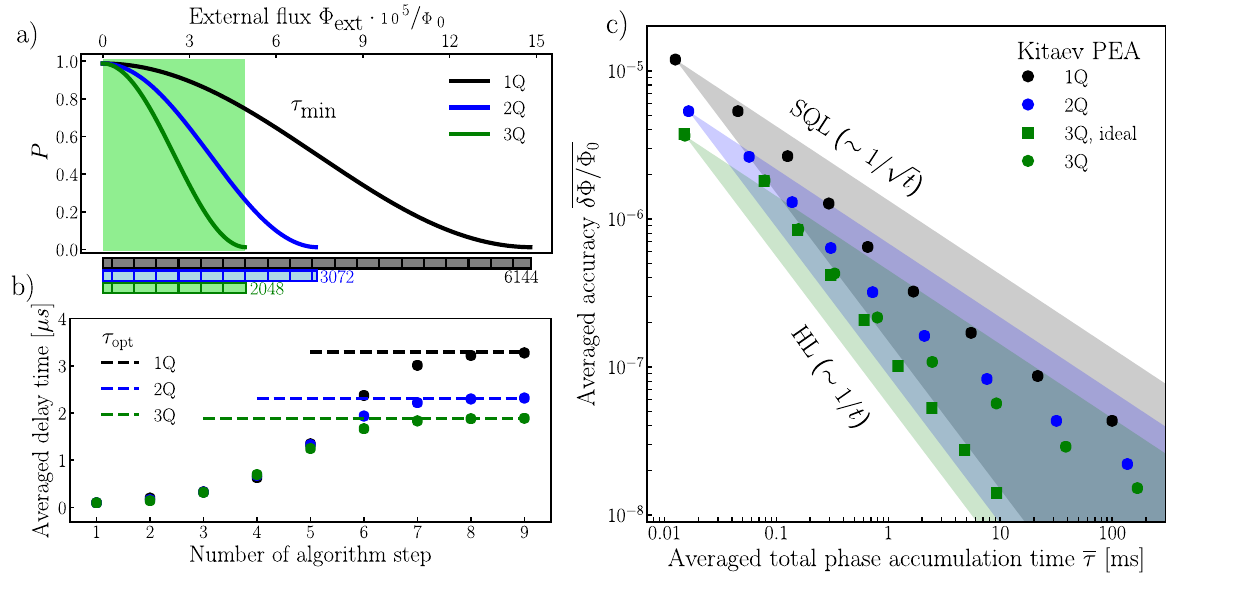}
\caption{(a) Calibration patterns at the minimal delay time used in the simulations for the single-qubit sensor and for the sensors with $2$ or $3$ entangled qubits. (b) Averaged delay times for different steps of the Kitaev PEA and sensors comprising the different number of qubits. (c) Comparison of the flux sensing accuracy scaling with the averaged phase accumulation time for the sensors comprising the different number of qubits. Limitations in coherence shift the accuracy scaling away from the Heisenberg (HL) to the Standard Quantum limit (SQL), where $t$ is the sensing time.}
\label{Fig6}
\end{figure}
The number of flux values in the calibration pattern for the $3$-qubit case, 2048, is chosen to allow for 11 steps of the algorithm. Because the number of possible flux values is reduced by a factor of 2 after each step, the algorithm will choose which of the 2 remaining flux values is more probable at the last step for the 2048 initial values. We perform only 9 steps for the sake of time in our simulations. If the sensor is exposed to the measured field only during the phase accumulation time, the dynamic range of fluxes measured with $N$ entangled qubits is $\Phi_{\rm{max}}^N=\pi/(\tau_{\rm{min}}\vert d\omega_q/d\Phi\vert N)$, where $\tau_{\rm{min}}$ is the minimal time required for switching the external field on and off. Thus, for a sensor with $N$ entangled qubits, the dynamic range is reduced as $\sim\ 1/N$ in comparison with a single qubit sensor. With this in mind, all flux grids are chosen such that the grids for $2$ and $3$-qubit sensors form the subsets of the grid for the single-qubit sensor (Fig.~\ref{Fig6}(a)). We choose $F=256$ flux values to be measured from the flux grid of the $3$-qubit sensor so that it is also possible to measure them with the two other sensors. We repeat the algorithm $M = 24$ times at each of the $F = 256$ flux values. Fig.~\ref{Fig6}(b) shows the obtained delay times for every step of the algorithm averaged, first, over $M = 24$ repetitions and, then, over $F = 256$ flux values. One can see that the delay times grow on average from step to step, and tend towards the optimal delay times of the sensor $\tau_{\rm{opt}}^N$. In the general case of $N$ qubits, this time reads
\begin{equation}
\tau_{\rm{opt}}^N=\frac{-NA+\sqrt{N^2A^2+8B^2N}}{4B^2N}
\label{eq_22}
\end{equation}
which corresponds to Eq.~(\ref{eq_15}) for $N=1$. With the reduction of the coherence time and the optimal sensing time for the entangled systems ($N >1$), the delay times start to saturate at the earlier steps.

Fig.~\ref{Fig6}(c) shows the results of the simulations. We compute the phase accumulation time $\tau_{j,k,l}$ for every flux value ($j$), repetition ($k$), and the step ($l$), and then the averaged total phase accumulation time, $\overline{\tau_l}$, for every step as
\begin{equation}
\overline{\tau_l}=\frac{1}{F}\sum_{j=1}^F\frac{1}{M}\sum_{k=1}^M\tau_{j,k,l},\quad \tau_{j,k,l}=\sum_{i=1}^l\tau_i^{(j,k)}n_i^{(j,k)}.
\label{eq_23}
\end{equation}
Here, $\tau_i^{(j,k)}$ and $n_i^{(j,k)}$ are the delay time for the step number $i$ and the number of measurements done at this step for the $j$-th flux value in the $k$-th repetition, respectively. In our simulations, we use $\sigma_0=\sigma_1=1.0$ for the widths of the measurement outcome normal distributions for states $\vert 0\rangle$ and $\vert 1\rangle$, and $\epsilon=0.01\%$ for the error probability ~\cite{Danilin_magnetometry}. These determine the number of measurements $n_i$ done at each step and also the condition to terminate the step and discard less probable flux values. By the end of step $l$ of the algorithm, we have a probability distribution for every flux value $\Phi_j, j\in [1,F]$ chosen to be measured and every repetition $k\in [1,M]$. We use this distribution to compute the mean flux values $\hat{\Phi}_{jkl}$ and find the averaged flux accuracy for every step as
\begin{equation}
\overline{\left(\frac{\delta\Phi}{\Phi_0}\right)_l}=\sqrt{\frac{1}{\Phi_0^2F}\sum_{j=1}^F\frac{1}{M-1}\sum_{k=1}^M(\hat{\Phi}_{jkl}-\Phi_j)^2}.
\label{eq_24}
\end{equation}
Dependencies of the averaged flux accuracy $\overline{(\delta\Phi/\Phi_0)_l}$ on the averaged total phase accumulation time $\overline{\tau_l}$ are shown in
Fig.~\ref{Fig6}(c).

We compare the improvements of the flux sensing accuracy with the phase accumulation time for the sensors comprising a single qubit, and $2$ or $3$ entangled qubits, in Fig.~\ref{Fig6}. One can see that, even at the first step of the algorithm, sensors with $2$ or $3$ entangled qubits have an advantage in the flux accuracy due to the smaller dynamical range. The improvement in flux sensing accuracy from the first to second algorithm step varies for each system. In fact, the accuracy obtained from the $3$-qubit sensor at the second algorithm step is nearly the same as that obtained by the $2$-qubit sensor at this step. The calibration patterns at the minimal delay time $\tau_{\rm{min}}$ cover different ranges of probability values for the flux interval values from which we measure, see Fig.~\ref{Fig6}(a). The delay time does not always increase when we progress from the first to the second step of the algorithm for 3-qubit case, in contrast with 1 and 2-qubit cases. As a result, we spend a longer time on average at the second step of the algorithm for the sensor with $3$ qubits. This partial setback is made up for in the later steps of the algorithm. The scaling of the flux accuracy is close to the HL scaling for all considered sensors after the first two steps of the algorithm. When the averaged delay time approaches the optimal sensing time and starts to saturate (Fig.~\ref{Fig6}(b)), the accuracy
scaling deviates from the HL scaling and returns gradually back to the SQL scaling. The shorter the coherence time of the sensor, the shorter the optimal sensing time and the faster this transition happens. The sensors with $2$ and $3$ entangled qubits therefore deviate from the HL scaling at the earlier steps of the algorithm. Nevertheless, the accuracies at the same phase accumulation time achieved by the sensors with entangled qubits are always better than those of the sensor based on a single qubit. The sensor with $3$ entangled qubits also proves to be better than with $2$ entangled qubits. The advantage in the accuracy is reduced as the crossover from the HL scaling to the SQL scaling occurs, but the advantage from the earlier steps of the algorithm is not completely lost at the later steps. For the case of $3$ entangled qubits without decoherence, the scaling follows the HL ($\sim 1/t$) law, Fig.~\ref{Fig6}(c) green squares, and the accuracy reaches values of $\sim 10^{-8}\Phi_0$ in time interval order of magnitude shorter than for the case when decoherence is present. This demonstrates the potential of the method for sensors with improved coherence times. 

In practice, the calibration of a sensor employing PEA \- the measurement of the probability pattern $P_{\vert 10..0\rangle}$ \- can take a long time. To mitigate this, FPGA-based electronics and parametric amplifiers can be used for a single-shot readout and fast reset of the sensor qubits~\cite{FPGA_Gebauer}. If the duration of control pulses and the time to read out and reset the qubits are much shorter than the coherence time of the sensor, the total sensing time will almost entirely consist of the phase accumulation time. This will noticeably shorten the calibration and speed up the sensing itself.

\section{\label{sec:Conc} Conclusions}

The article proposes a way of using multi-qubit superconducting structures for sensing external static fields. Estimates of relaxation and dephasing rates are made for a given design of a frequency-tunable transmon qubit for its use as a quantum sensor. Based on these estimates, the optimal point for sensing in the flux bias is found for different maximal qubit transition frequencies. The optimal point for sensing provides the maximal sensitivity to the external fluxes. A strategy involving the Kitaev phase estimation algorithm run on the $2$ and $3$-qubit entangled states for sensing and metrology with superconducting quantum circuits is considered with the aim of going beyond the SQL scaling in the time domain. This is applied at the optimal sensing point. The advantages in sensing accuracy that come from the increased speed of the phase accumulation for entangled states of $N$ qubits are demonstrated in simulation. The simulation results show that the sensing accuracy of the magnetic flux threading the SQUID-loops of the sensor qubits  reaches the values of $10^{-8}\Phi_0$ in a relatively short measurement time \--- outperforming the SQL scaling. The accuracy improvement from one step of the algorithm to the next one follows the HL scaling for the initial steps of the sensing algorithm. Sensors with more coherent qubits will follow the HL scaling for more steps and achieve better final accuracy in shorter sensing time.  

\ack{The authors are thankful for the support from the European Research Council (ERC) under the Grant Agreement No. 648011, the Engineering and Physical Sciences Research Council (EPSRC) under the Grant Agreements No. EP/T018984/1 and No. EP/T001062/1.}

\appendix
\section{}\label{A1}

\begin{figure}[!h]
\centering
\includegraphics[width=\columnwidth]{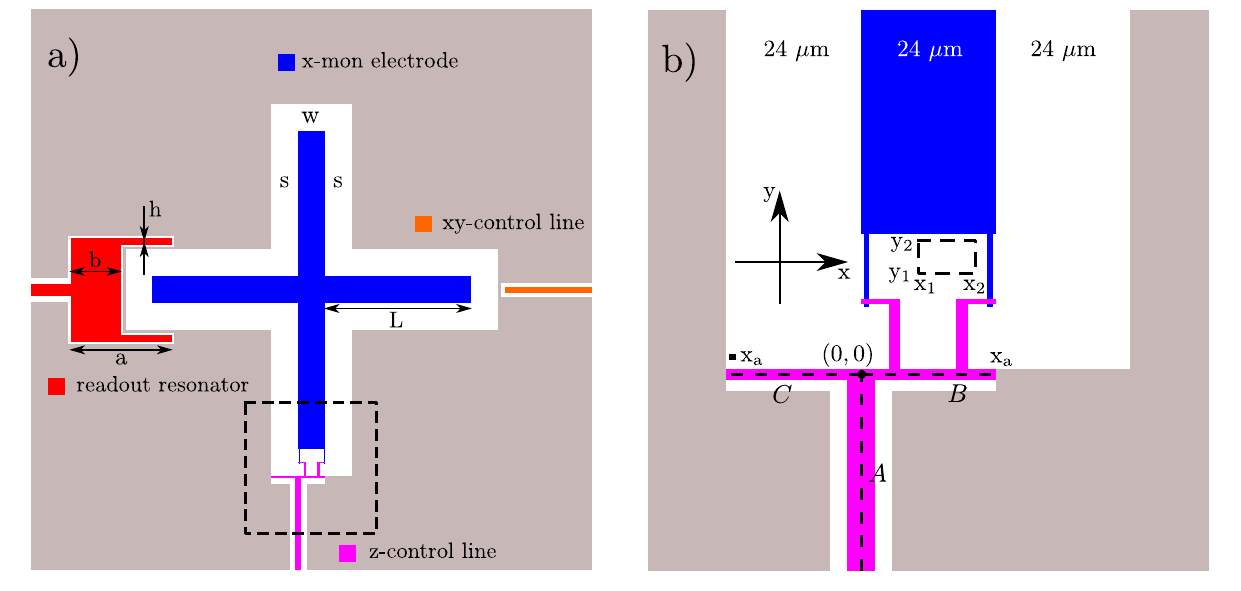}
\caption{(a) X-mon type superconducting qubit design. xy- and z-control lines as well as the end of the readout resonator are shown in the figure in different colors. The Maxwell capacitance matrix was computed numerically in Comsol Multiphysics to find the circuit parameters. Dashed rectangle denotes the area shown in (b). (b) Sketch of the flux bias line and the SQUID loop used in the analytical computation of the mutual inductances of the circuit.}
\label{FigS1_v1}
\end{figure}

The estimation of the circuit parameters is done for the transmon qubit design shown in Fig.~\ref{FigS1_v1}. This is the x-mon type~\cite{x-mon} of transmon qubit. The SQUID loop which makes the qubit tunable is at the end of one arm of the x-mon qubit. The cross-shaped x-mon electrode has arm width $w=24\ \mu\rm{m}$, arm length $L = 130\ \mu\rm{m}$, and an x-mon electrode to ground plane gap of $s = 24\ \mu\rm{m}$. The coupler to the readout resonator and the xy- and z-control lines are shown in different colours in the figure. The geometric parameters of the coupler to the readout resonator are following: $a=90\ \mu\rm{m},\ b=45\ \mu\rm{m},\ \rm{and}\ h=6\ \mu\rm{m}$, see Fig.~\ref{FigS1_v1}(a). The width of the ground plane line separating the coupler from the x-mon electrode is $2\ \mu\rm{m}$. We perform numerical simulation of the Maxwell capacitance matrix in Comsol Multiphysics to estimate the capacitances between the x-mon electrode and the ground $C_{qg}\simeq 76\ \rm{fF}$, the x-mon electrode and the resonator $C_{qr}\simeq2\ \rm{fF}$, and the x-mon electrode and the xy-control line $C_{c}\simeq 0.2\ \rm{fF}$. This can be done by setting the potential of either the x-mon electrode, resonator coupler electrode or the xy-control line to $1\ \rm{V}$, keeping the potentials of all other electrodes at $0\ \rm{V}$, and solving for the charge required to achieve this potential difference. It is also possible to find the voltage divider $\beta\simeq 0.03$~\cite{Koch_2007} required for the coupling strength estimation, Eq.~(\ref{eq_3}). For this, we set the potential of the xy-control line to $1\ \rm{V}$, keep the charge of the x-mon electrode at $0\ \rm{C}$ and the ground plane voltage at $0\ \rm{V}$, and solve for the potential of the x-mon electrode. The value of this potential is numerically equal to the divider $\beta$. With the use of these circuit parameters, we find the x-mon charging energy $E_C/h\simeq 0.254\ \rm{GHz}$, where $h$ is the Planck constant. These parameters also enter all decoherence rates we estimate in Sec.~\ref{sec:Sec2}.

To estimate the inductances between the flux bias line and the qubit SQUID loop $M$ and between the flux bias line and the total circuit of the qubit, we employ the Biot-Savart-Laplace law to analytically compute the magnetic field magnitudes \begin{equation}
\vec{B}(x,y) = \frac{\mu_0}{4\pi}\int_l\frac{Id\vec{l}\times\vec{r}^\prime}{\vert\vec{r}^\prime\vert^3} 
\label{eq_S1}
\end{equation}
for different points in the plane of the circuit produced by the current $I$ flowing in the bias line, see Fig.~\ref{FigS1_v1}(b). Here, $\mu_0$ is the free space permeability, $d\vec{l}$ is the current element on the chosen path $l$, and $\vec{r}^\prime$ is the vector pointing from the element $d\vec{l}$ to the point $(x,y)$. We subdivide the flux bias line into 3 parts A, B, and C and compute the fields produced by the currents in each of the parts separately. The currents in the parts B and C are half as large as in the part A. Because the current is spread into a broad angle after it reaches the ends of the parts B and C and enters the ground plane, we do not consider the magnetic field caused by the currents in the ground plane. The length of the flux bias line arm is $x_a=24\ \mu\rm{m}$. The width of the part A of the flux bias line is $5\ \mu\rm{m}$, and the widths of the parts B and C are $2\ \mu\rm{m}$. After performing the integration for a point $(x,y)$ (Eq.~\ref{eq_S1}) as it is shown in Fig.~\ref{FigS1_v1}(b), the magnitudes of the magnetic field can be expressed as
\begin{equation}
B(x,y)=
\begin{cases}
\frac{\mu_0I}{4\pi}\left(\frac{1}{x}-\frac{y}{x\sqrt{x^2+y^2}}\right), & \rm{for}\ A\\
\frac{\mu_0I}{8\pi}\left(\frac{x}{y\sqrt{x^2+y^2}}-\frac{(x-x_a)}{y\sqrt{(x-x_a)^2+y^2}}\right), & \rm{for}\ B\\
\frac{\mu_0I}{8\pi}\left(\frac{(x+x_a)}{y\sqrt{(x+x_a)^2+y^2}}-\frac{x}{y\sqrt{x^2+y^2}}\right), & \rm{for}\ C.
\end{cases}
\label{eq_S2}
\end{equation}

We can compute the magnetic flux threading the SQUID loop by integrating these fields over the area of the loop
\begin{equation}
\Phi = \int\limits_{x_1}^{x_2}\int\limits_{y_1}^{y_2}B(x,y)dxdy.
\label{eq_S3}
\end{equation}
The area of the SQUID loop consists of two rectangles with areas $125\ \mu\rm{m}^2$ and $241.5\ \mu\rm{m}^2$ (see Fig.~(\ref{FigS1_v1})(b)). The coefficient in front of the current $I$ gives the mutual inductance between the flux bias line and the SQUID loop $M=2.08\ \rm{pH}$, $\Phi=MI$. This value of inductance corresponds to $1\ \rm{mA}$ periodicity in the bias current of the qubit spectrum. While making this computation, we do not take into account the expulsion of the magnetic field from the superconducting film of the circuit, and this assumption will be confirmed in the experiment.

By applying the integration to the area of the entire gap between the x-mon electrode and the ground plane, we find the parasitic inductance $M^\prime=0.22\ \rm{pH}$. Its non-zero value arises from the asymmetric position of the flux bias line relative to the x-mon arm. Both inductances, $M$ and $M^\prime$, enter the relaxation rate $\Gamma_1^{\rm{ind}}$ in Eq.~(\ref{inductive_coupling_rate}).

\bibliographystyle{unsrt}
\nocite{*}
\bibliography{references_v1}

\end{document}